\documentclass[pre,amsmath,amssymb,twocolumn,nofootinbib,floatfix]{revtex4-2}
\usepackage[colorlinks=true,linkcolor=blue,urlcolor=blue,citecolor=blue]{hyperref}
\usepackage{graphicx}\usepackage{bm,color}
\usepackage{color}
\usepackage{appendix}
\usepackage{times}

\definecolor{orange}{rgb}{1,0.6, 0}
\definecolor{darkergreen}{rgb}{0,0.7,0}
\definecolor{grey}{rgb}{.4,.4,0.4}

\usepackage{float}
\usepackage[caption = false]{subfig}

\definecolor{labelkey}{cmyk}{.4,.2,0,0}

\renewcommand{\doi}[2]{\href{http://dx.doi.org/#1}{#2}}
\newcommand{\arxiv}[1]{\href{http://arxiv.org/abs/#1}{#1}}
\newcommand{\link}[2]{\href{http://#1}{#2}}

\newcommand{\Eq}[1]{Eq.~(\ref{#1})}
\newcommand{\Eqs}[1]{Eqs.~(\ref{#1})}
\newcommand{\eq}[1]{(\ref{#1})}

\newcommand{\blue}{\color{blue}}
\definecolor{dg}{rgb}{0.0, 0.5, 0.0}

\newcommand{\bea}{\begin{eqnarray}}
\newcommand{\eea}{\end{eqnarray}}
\newcommand{\beq}{\begin{equation}}
\newcommand{\eeq}{\end{equation}}
\newcommand{\red}{\color{red}}

\newcommand{\rme}{\mathrm{e}}
\newcommand{\rmd}{\mathrm{d}}
\newcommand{\nn}{\nonumber}
\renewcommand{\epsilon}{\varepsilon}
\newcommand{\nott}[1]{}
\newcommand{\ca}[1]{{\cal #1}}
\newcommand{\be}{\begin{equation}}
\newcommand{\ee}{\end{equation}}

\newlength{\bilderlength}

\graphicspath{{figures/},{}}

\renewcommand{\paragraph}{\subsubsection*}

\begin{document}


\title{Experimental test of Sinai's model  in DNA unzipping}
\author{Cathelijne ter Burg${^{1}}$,  Paolo  Rissone${^{2}}$, Marc Rico-Pasto${^2}$, Felix Ritort${^{2,3}}$ and Kay J\"org Wiese${^1}$}
  \affiliation{\mbox{{${}^1$}Laboratoire de Physique de l'E\'cole Normale Sup\'erieure, ENS, Universit\'e PSL, CNRS, Sorbonne Universit\'e,} \mbox{Universit\'e Paris-Diderot, Sorbonne Paris Cit\'e, 24 rue Lhomond, 75005 Paris, France.}
  \mbox{{${}^2$}Small Biosystems Lab, Condensed Matter Physics Department, Universitat de Barcelona,} \mbox{Carrer de Mart\'i i Franqu\`{e}s 1, 08028 Barcelona, Spain.}
  \mbox{{${}^3$}Institut de Nanoci\`encia i Nanotecnologia (IN2UB), Universitat de Barcelona, 08028 Barcelona, Spain}}

\begin{abstract}
The experimental measurement of correlation functions and critical exponents in disordered systems is key to testing renormalization group (RG) predictions. We mechanically unzip single DNA hairpins 
with optical tweezers, an experimental realization of the diffusive motion of a particle in a one-dimensional random force field, known as the {\em Sinai model}. We measure the unzipping forces $F_w$ as a function of the trap position $w$ in equilibrium and calculate the force-force correlator $\Delta_m(w)$, its amplitude, and correlation length, finding agreement with theoretical predictions.   We study the universal scaling properties since the effective trap stiffness $m^2$ decreases upon unzipping. Fluctuations of the position of the base pair at the unzipping junction $u$ scales as $u \sim m^{-\zeta}$, with  a {\em roughness exponent} $ \zeta=1.34\pm0.06$,   in   agreement with the analytical prediction $\zeta = \frac{4}{3}$. Our study provides a single-molecule test of the functional RG approach for disordered elastic systems in equilibrium.
\end{abstract} 

\maketitle

\noindent{\it Introduction.} Heterogeneity and disorder pervade physical and biological matter \cite{bouchaud1990anomalous,kardar1998nonequilibrium,kirkpatrick2015colloquium}. Since Schr\"odinger's conception of the gene as an a-periodic crystal \cite{schrodinger1944life}, 
disorder is recognised as a crucial ingredient for life \cite{varn2016did}. 
The readout of the genetic information encoded in DNA can be modeled with polymers in random potentials, such as  Sinai's model \cite{Sinai1983}. The latter describes the dynamics of a particle diffusing in a one-dimensional random-force field, a  suitable model for the mechanical unzipping of the DNA double helix into single strands. 
Sinai's model is a special case ($d=0$) of  the universal field theory of disordered elastic systems in $d$ dimensions, where one  can analytically  calculate force correlations.
The latter were measured in  contact-line depinning ($d = 1$) \cite{LeDoussalWieseMoulinetRolley2009}, Barkhausen noise ($d = 2$) \cite{terBurgBohnDurinSommerWiese2021} and RNA-DNA peeling ($d = 0$) \cite{WieseBercyMelkonyanBizebard2019}. While   these experiments are for depinning, i.e.~nonequilibrium, an experimental test of the equilibrium universality class is  lacking. Here we test   universality of equilibrium-force correlations as predicted by Sinai's model in DNA unzipping.  The model parameters are naturally changed during the experiment allowing us to monitor the functional RG flow. 
 
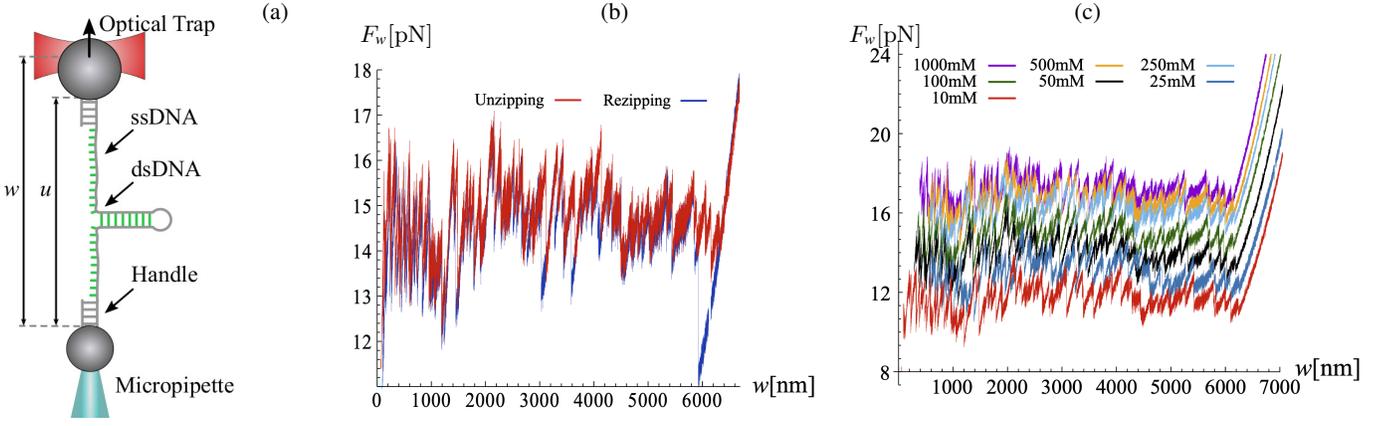
\begin{figure*}[t]
{\setlength{\unitlength}{1cm}\begin{picture}(17.8,5.5)
\put(0.0,0){ \includegraphics[width=.17\textwidth]{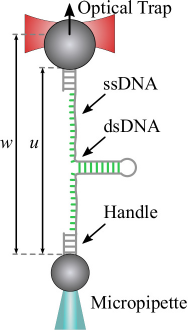}}
\put(4.4,0.0){ \includegraphics[width=.38\textwidth]{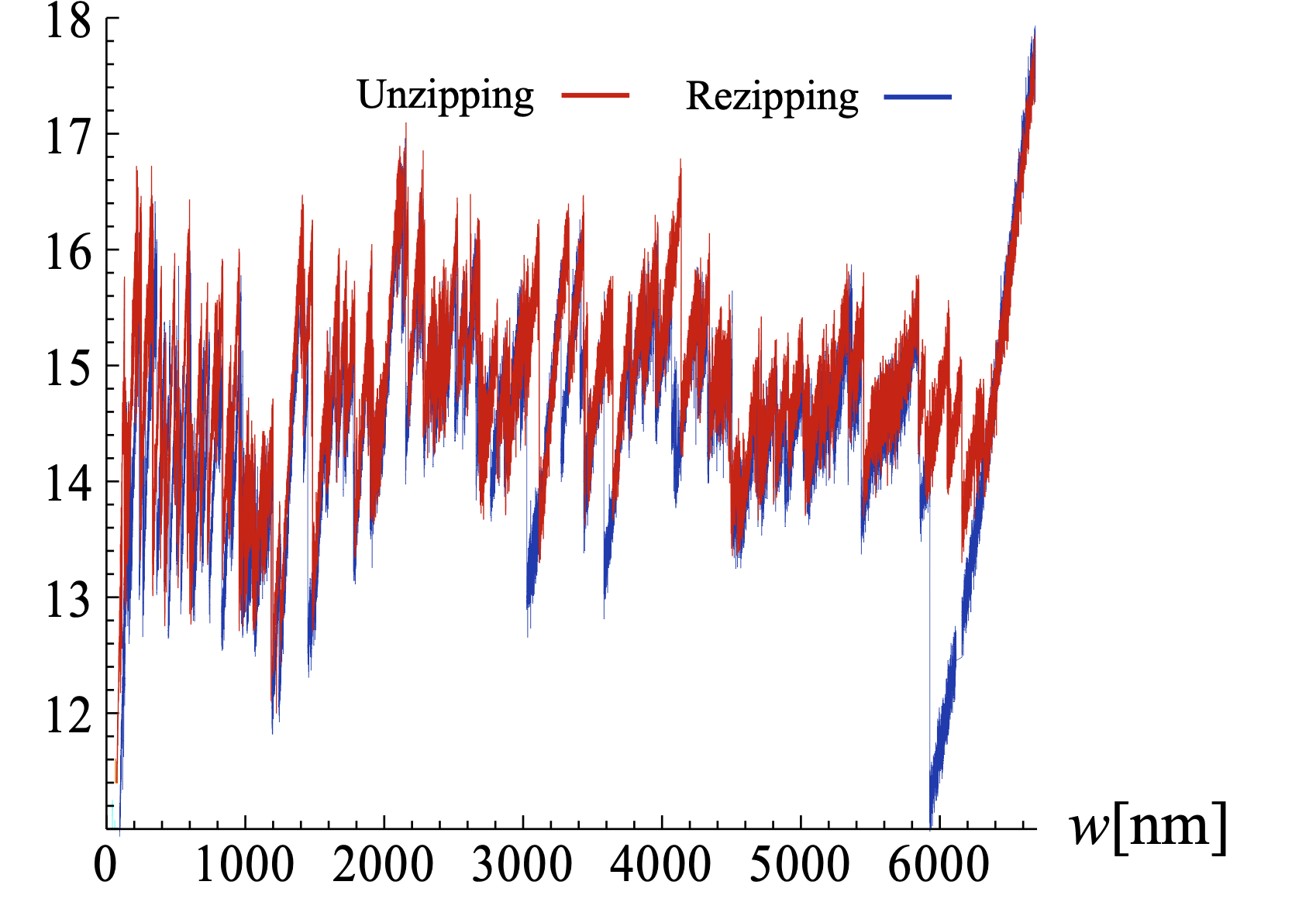}}
\put(11.3,0.1){ \includegraphics[width=.38\textwidth]{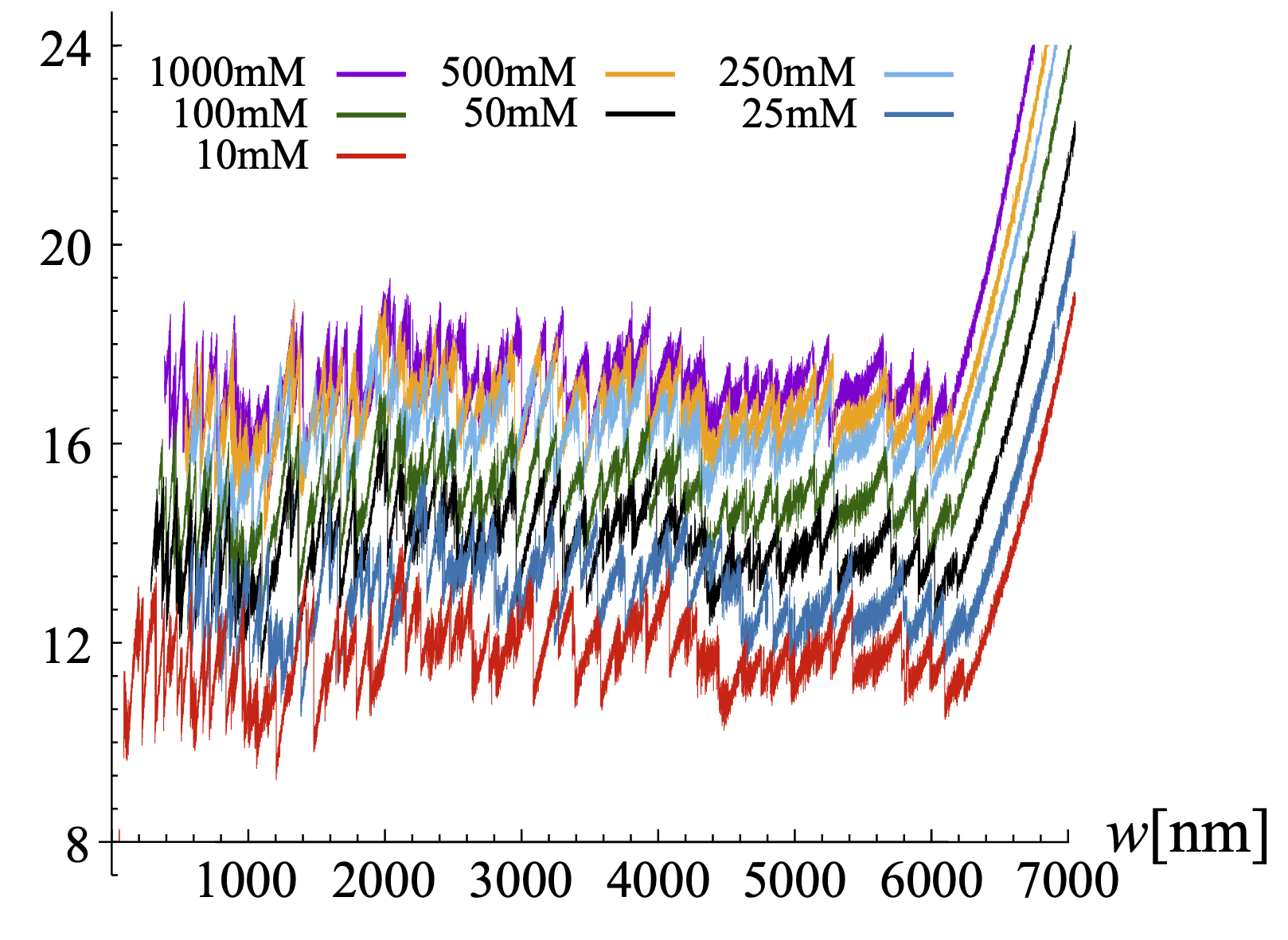}}
\put(3.5 ,5.3){(a)}
\put(8.0 ,5.3){(b)}
\put(4.8 ,5.0){$F_{\emph w} [{\rm pN} ]$}
\put(14.3 ,5.3){(c)}
\put(11.3 ,5.0){$F_{ \emph w} [{\rm pN} ]$}
\end{picture}}
\caption{(a) Experimental setup. (b) Unzipping (red) and rezipping (blue) FDC's demonstrating equilibrium behaviour. The residual hysteresis at the end of the FDC  is due to the DNA end-loop that slows down the initiation of stem formation upon reconvolution. (c) Experimental FDC's, ${F_w}$, for various salt concentrations. The  mean pinning force varies between $12$-$17 {\rm pN}$, and is  non-universal.}
\label{f:Fig1}
\end{figure*}

In the experiment, a DNA hairpin of 6.8k base pairs (BPs) is held between two beads. One is fixed at the tip of a micropipette, the other is optically trapped (Fig.~\ref{f:Fig1}(a) and  Supp.~Mat.~Sec.~\ref{SI:Experiment}). By moving the optical trap at a speed $v \approx $ 10nm/s, the double-stranded DNA (dsDNA) is mechanically pulled and converted into two single strands (ssDNA). The measured force-distance curve (FDC) shows a sawtooth pattern characteristic of stick-slip dynamics (Fig.~\ref{f:Fig1}(b), red curve). The hairpin unzips at a critical mean pinning force $f_{\rm c}\approx 15{\rm pN}$, fluctuating in the range 12-17pN. 
Once the hairpin is unzipped, the reverse process starts (Fig.~\ref{f:Fig1}(b), blue curve): the optical trap moves backward and the hairpin refolds into the dsDNA native conformation.  The absence of hysteresis between rezipping and unzipping FDCs and the fact that there is a single reaction coordinate, implies that the system is in equilibrium. 

During unzipping, the base pair at the junction separating dsDNA from ssDNA is subject to random forces generated by the neighbouring monomers, and modeled by the motion of a single particle ($d=0$) in a random potential that belongs to Sinai's universality class \cite{Sinai1983}.   The number of unzipped BPs is a well-defined reaction coordinate. Opening (closing) one BP can be seen as a particle hopping to the right (left).  
We changed salt concentration from 10mM to 1000mM NaCl, Fig.~\ref{f:Fig1}(c), modulating the strength of BP interactions.

\medskip
\noindent{\it The Model.} 
The motion of the  base pair at the junction can be modeled by a Langevin equation (see Supp.~Mat.~Sec.~\ref{SupplEffTheory} for  the  derivation) 
\be
\label{eq:EqMotion}
 \frac{\partial u}{\partial t} =  m^2 (w - u) + F(u) +\eta_u(t)\, ,
\ee
where $u(t)$ is the extension of the molecular construct, $w$ the relative trap-pipette position (Fig.~\ref{f:Fig1}(a)), and $m^2$ the effective stiffness of the 
molecular construct. The random force is $F(u) = - V'(u)$, where $V(u)$ is the free energy stored in the partially hybridized hairpin. $F(u)$ acts at the hairpin junction and is determined by hydrogen bonding and stacking interactions between consecutive base pairs. Using the nearest-neighbour model one can show that  these forces are random, and that their distribution  is roughly a Gaussian (Supp.~Mat.~Sec.~\ref{s:Distribution of binding energies}). In equilibrium, $\frac{\partial u}{\partial t}  \approx 0$, so the force $F(u)$ applied to the hairpin in \Eq{eq:EqMotion} is counteracted by the force $F_w$ exerted on the bead by the optical trap.
For a fixed trap position $w$, $F_w$ and $u$ fluctuate due to the thermal noise and the BP breathing dynamics. The equilibrium force correlations are defined as,
\be 
\Delta_{m,T}(w - w^\prime)   =    \overline{F_w F_{w^\prime}}^{c}  =   \overline{F_w F_{w^\prime}} -   \overline{F_w  } \; \overline{ F_{w^\prime}} ,  \label{Correlations}
\ee
where $\overline{(\dots)}$ stands for a double thermal and disorder average. Correlations depend on the value of  $m^2$, through the $m$-dependence in \Eq{eq:EqMotion}. They also depend on temperature $T$, which leads to a rounding of $\Delta_{m,T}(w)$ at small $w$ (see below).

The FDCs in Figs.~\ref{f:Fig1}(b) and (c) show a sawtooth pattern characterized by segments of increasing force $F_w$, followed by abrupt drops caused by the  cooperative unzipping of groups of base pairs in the range of 10-100 basepairs \cite{HuguetFornsRitort2009}.

The slope of each segment, equivalent to the effective stiffness $m^2$, decreases with $w$, permitting us to measure the scaling of $\Delta_{m,  T}(w)$ with $m^2$. In fact, $m^2$ depends on the combined effects of the optical trap, and the elastic response of the molecular construct (ssDNA and dsDNA handles). It  can be written as (see \Eq{eq:eff_stiffness2})
\be
\label{EffMassFormula}
\frac{1}{m^2} = \frac{1}{k_{\rm b}} +\frac{w}{\overline{z_1} k_1}, 
\ee
with $ k_{\rm b}$ the trap stiffness, and $\overline{z_1},k_1$ the mean extension and stiffness of one nucleotide at the unzipping force. Modeling the elastic response of the hairpin \cite{AlemanyRitort2014} shows that  $k_1 \approx 130 {\rm pN/ nm}$ and $\overline{z_1} \approx 0.45 {\rm nm}$ at the unzipping force $f_{\rm c} \approx 15$ pN, which gives a slope of about $(\overline{z_1}k_1)^{-1} \approx 0.02 {\rm pN}^{-1}$. 
\begin{figure}[b]
\fboxsep0mm
\setlength{\unitlength}{1cm}
{\begin{picture}(8.6,6.5)
\put(0,0){\includegraphics[width=8.6cm]{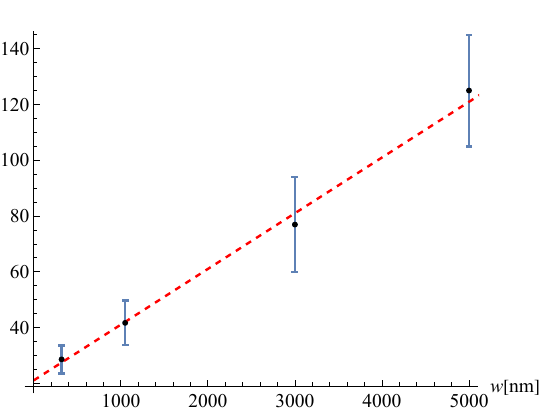}}
\put(0.85,3.25){\includegraphics[width=4.4cm]{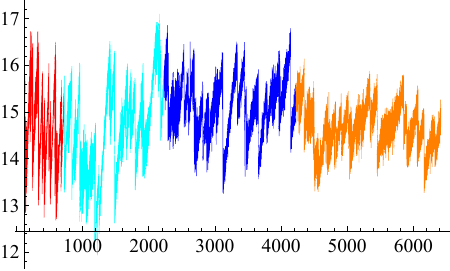}}
\put(-0.5,6.3){ \color{black} $m^{-2}[{\rm pN}^{-1}\cdot{\rm nm}]$ }
\end{picture}}
\caption{Variation of the effective stiffness $m^2$ versus $w$ according to \Eq{EffMassFormula}. The points correspond to the measured values of $1/m^2$ for the four FDC regions (each one shown with a different colour in the inset). The fit to data (dashed line) and the extrapolation to $w=0$ gives the stiffness of the optical trap, $k_{\rm b} = 0.05 \pm 0.01 { \rm pN \cdot nm^{-1} }$. The inset illustrates the four studied regions in a FDC at 1M NaCl.} 
\label{f:Fig2}
\end{figure}
\Eq{EffMassFormula} implies that the larger the length of the unpaired DNA, the lower the effective stiffness  $m^2$. To verify this, we  split the FDCs into four regions (inset of Fig.~\ref{f:Fig2}). While  smaller regions   have  smaller variations in $m^2$, regions must be taken sufficiently large for a reliable statistics. \Eq{EffMassFormula} agrees with the experimental data shown in Fig.~\ref{f:Fig2}. 

Force correlations in Sinai's model can be framed in terms of the functional renormalisation group (FRG).
The FRG arises as the field theory of disordered systems for interfaces 
\cite{Wiese2021,BalentsLeDoussal2004,ChauveGiamarchiLeDoussal2000,DSFisher1986,NattermannStepanowTangLeschhorn1992,NarayanDSFisher1993a,ChauveLeDoussalWiese2000a,LeDoussalWieseChauve2002,LeDoussalWieseChauve2003,le2006can,WieseHusemannLeDoussal2018,HusemannWiese2017,zinati2022stochastic}, generalising the $d = 0$ case described by the Sinai model. The FRG predicts two universality classes, critical depinning (non-equilibrium) and equilibrium (considered here).
In equilibrium, the  $T\to 0$ limit of $\Delta_{m, T}(w) $ in \Eq{Correlations}, can be written as
\be \label{DeltaMDef}
\Delta_m(w) = m^4 \rho_m^2 \tilde \Delta(w/\rho_m) ,\qquad \rho_m\sim m^{-\zeta},
\ee 
with   $ \tilde \Delta(\sf w) $ the shape function, $\zeta$ the roughness exponent, and   ${\sf w}=w/\rho_m$ the rescaled dimensionless distance. The FRG allows for observables to be computed perturbatively in an expansion around the upper critical dimension,  parameterised by $\epsilon=4-d$. The shape function    $ \tilde \Delta(\sf w) $ is the fixed point of the FRG flow equation
 \be
0 =   (\epsilon  {-} 2 \zeta) \tilde{\Delta}({\sf w}) + \zeta {\sf w} \tilde{\Delta}^\prime({\sf w})  
  -   \frac{1}{2}  \partial_{{\sf w}}^2\bigl[ \tilde{\Delta}({\sf w}) {-} \tilde{\Delta}(0) \bigl] ^2   
+ \dots  \label{1loopFRG}
\ee
The dots represent higher-loop corrections in $\epsilon$, currently known up to 3-loop order \cite{ChauveLeDoussalWiese2000a,LeDoussalWieseChauve2002,LeDoussalWieseChauve2003,WieseHusemannLeDoussal2018,HusemannWiese2017,terBurgWiese2022}. For the equilibrium random-field, $\zeta = (4-d)/3 $, which gives $\zeta = 4/3$ for $d = 0$. This result is derived by integrating \Eq{1loopFRG} from $\sf w=0$ to $\sf w=\infty$. It is  exact to all orders in the loop expansion.
\Eq{1loopFRG} 
predicts that   $ \tilde \Delta(\sf w) $ has a cusp  at $\sf w = 0$ which is rounded   at finite $T$. 
Generalization of the FRG equation \eqref{1loopFRG} to finite $T$ allows us to estimate the size of the   rounded region. An explicit relation between $\Delta_m(w)$ and $\Delta_{m,T}(w)$ was derived in \cite{BalentsLeDoussal2004,ChauveGiamarchiLeDoussal2000,Wiese2021}, 
\be
\Delta_{m,T}(w)   \approx   {\cal N}   \Delta_m(\sqrt{w^2 + t^2})\label{BdrLayer}, \qquad  
t = 
 \frac{6 m^2 k_{\rm B}T  }{\epsilon |\Delta^\prime_m(0)|}  . 
\ee
It has been shown that the RG flow \eq{1loopFRG} preserves the area under  $\Delta_{m,T}(w)$ for all $T$ \cite{terBurgWiese2022}. Therefore, we can use the measured $\Delta_{m,T}(w)$ and \Eq{BdrLayer} to determine the normalization factor $\ca N$ and $\Delta_m(w)$. Details about the procedure are given in Supp.~Mat.~Sec.~\ref{suppUnfolding}.


For the Sinai model, the shape function $\tilde \Delta$ in \Eq{1loopFRG} is known  analytically  \cite{LeDoussal2009,Wiese2021}, 
\bea
\tilde{\Delta} ({\sf w}) &= & - \frac{\rme^{-\frac{{\sf w}^3}{12}}}{4 \pi^{\frac{3}{2}}\sqrt{{\sf w}}} \int_{-\infty}^{\infty} {\rm d} \lambda_1 \int_{-\infty}^{\infty} {\rm d} \lambda_2 \rme^{-\frac{(\lambda_1 - \lambda_2)^2}{4{\sf w}}} \notag \\ 
& \times & \rme^{i\frac{{\sf w}}{2}(\lambda_1 + \lambda_2 )} \frac{\rm{Ai}^{\prime}(i \lambda_1)}{\rm{Ai}(i \lambda_1)^2}
\frac{\rm{Ai}^{\prime}(i \lambda_2)}{\rm{Ai}(i \lambda_2)^2} \notag\\ 
&\times & \biggl[ 1 + 2{\sf w} \frac{\int_{0}^{\infty}{\rm d}V\rme^{\sf w V}\rm{Ai}(i \lambda_1 {+}V) \rm{Ai}(i \lambda_2 {+}V)}{\rm{Ai}(i \lambda_1) \rm{Ai}(i \lambda_2)} \biggl].   ~~~~~~
\label{DeltaSinai}
\eea
Here $\rm {Ai}$ is the Airy function, and  
 $\zeta = 4/3 $ as in FRG. 
%
%

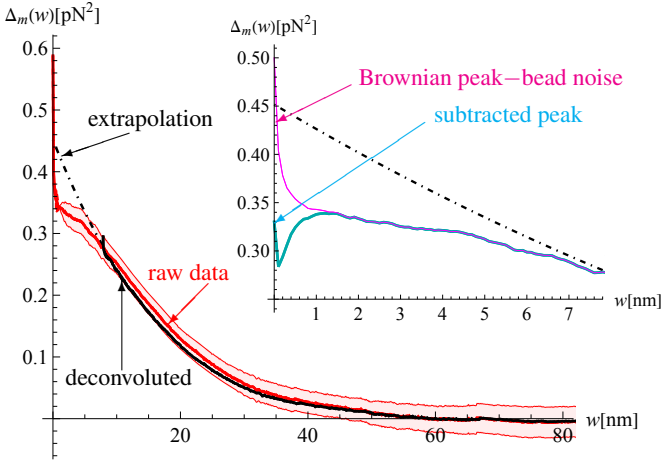
\begin{figure}[t]
{\setlength{\unitlength}{1cm}\begin{picture}(8.5,6.1)
\put(-.1,0){\includegraphics[width=.47\textwidth]{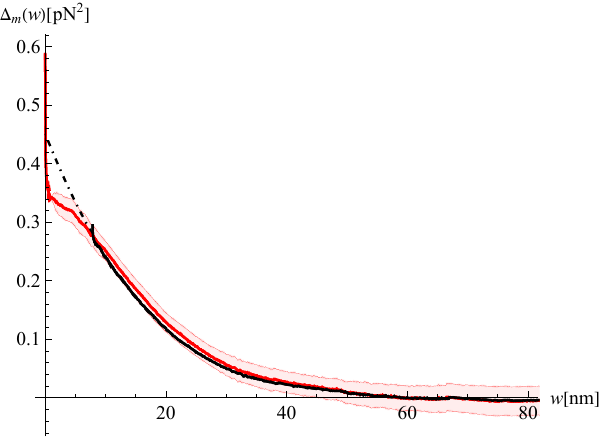}}
\put(2.9,1.8){\includegraphics[width=.32\textwidth]{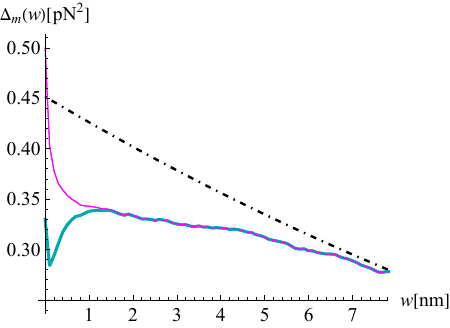}}
\put(0.7,1.){\color{black} deconvoluted}
\put(1.45,1.2){\color{black} \vector(0,5){1.2}}
\put(1.8,2.4){\color{red}raw data}
\put(4.6,5){\color{magenta}Brownian peak$-$bead noise}
\put(4.5,5){\color{magenta}\vector(-2,-1){1.0}}
\put(5.6,4.5){\color{cyan}subtracted peak}
\put(5.48,4.5){\color{cyan}\vector(-3,-2){2.0}}
\put(2.6,2.35){\color{red}\vector(-1,-1){0.53}}
\put(1.0,4.45){\color{black} {extrapolation}}
\put(1.55,4.35){\color{black}\vector(-3,-1){0.9}}
\end{picture}}
\caption{Measured $\Delta_{m,T}(w)$ for the first region (red). 1-$\sigma$ error is shown as a pink strip.
Deconvolution (black solid) and 
extrapolation to $w = 0$   (black dot-dashed). The inset shows   $\Delta_{m,T}(w)$ at short range with   subtraction of the peak at $w = 0$, as explained in the main text. 
}
\label{f:Fig3}
\end{figure}

\medskip
\noindent{\it Data analysis.}
We analysed 33 FDCs obtained by unzipping a 6.8kBP DNA hairpin in a broad range of salt conditions from 10mM to 1000mM NaCl  at $T=298 {\rm K}$. 
As illustrated in Fig.~\ref{f:Fig2}, we divided each FDC into four regions measuring the force correlations \eq{Correlations} for each region. 
Force correlations are equal within the experimental resolution for all salt conditions, as shown in Supplementary Fig.~\ref{f:suppsalt}. Although the effective stiffness of the molecular construct $m^2$ changes with salt, it changes much less than it does over the different unzipping regions for a fixed salt condition. To enlarge statistics we  averaged $\Delta_{m,T}(w)$ over all salts. Results for the first region are shown in Fig.~\ref{f:Fig3} (red line with red strip for error bars). 

To recover $\Delta_{m,T}(w)$ in \Eq{BdrLayer} we must subtract two sources of thermal noise, which are visible as a short-range correlated  peak at $w\approx 0$: Brownian fluctuations of the bead; and the breathing dynamics (opening and closing) of the DNA base pairs at the junction. 
First, bead-noise subtraction reduces the peak's amplitude $\Delta_{m,T}(w=0)$ from $\approx 0.6\rm pN^2$ (red in main plot of Fig.~\ref{f:Fig3}) to $\approx 0.5\rm pN^2$ (magenta line in the inset). Second, we estimated the effect of the breathing dynamics from numerical simulations of Sinai's model \cite{terBurgWiese2022}. This reduces the peak from $\approx 0.5\rm pN^2$ to $\approx 0.35\rm pN^2$ with a dip of amplitude $\approx 0.3\rm pN^2$ for $w<1$nm (cyan curve in the inset). This dip is also seen in simulations \cite{terBurgWiese2022}.
 From $\Delta_{m,T}(w)$ we derive the $T=0$ force correlations, $\Delta_m(w)$, by plotting the experimental data versus $\sqrt{w^2 + t^2}$, see \Eq{BdrLayer}, with $t$ given there ($T=298$K, $\epsilon=4$, $m^2$ from Fig.  \ref{f:Fig2}). We initially estimate $\Delta_m'(0)$ by  extrapolation of the raw data.
This gives $\Delta_{m}(w)$ for $w>t\approx 7\rm nm$ (black continuous line in Fig.~\ref{f:Fig3}). The extrapolated $\Delta_m(w)$ for $w<t$ (dot-dashed region) is obtained by fitting a second-order polynomial (black dot-dashed line in Fig. \ref{f:Fig3}). The whole procedure is iterated until convergence of $\Delta_m(w)$ is reached. 
As a consistency check we used the $T=0$ theory prediction $\Delta_m(w)$ together with \Eq{BdrLayer} to calculate $\Delta_{m,T}(w)$ for all regions, see Supp.~Fig. \ref{Fig8}. 


Force correlations in \Eq{BdrLayer} are described by three parameters: the correlation length  $\rho_m$ in the $w$ direction, the stiffness $m^2$ of the molecular construct, and the temperature $T$. With the measured value of $m^2$ (Fig.~\ref{f:Fig2}) and $k_{\rm B} T = 4.11 {\rm pN \cdot nm}$ we  use \Eq{BdrLayer} to predict $t$ ($\epsilon=4$ and $\Delta_m'(0)$ obtained from the small-$w$ extrapolation in Fig.~\ref{f:Fig3}). According to \Eqs{DeltaMDef} and \eq{BdrLayer}, the scale $\rho_m$ is the only fitting parameter, which we report on the table in Fig.~\ref{f:Fig5} for all four regions. Its value increases with $w$ indicating that FDCs become progressively less rough as unzipping progresses: For the first region, $\rho_m = 26.8 \rm nm$, which corresponds to 33 basepairs \cite{AlemanyRitort2014}, the typical size of 
avalanches that can be resolved in the FDC at the beginning of the unzipping process.

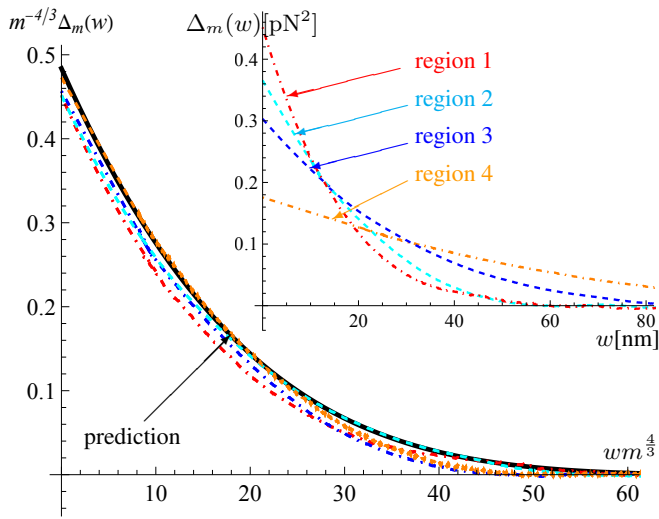
\begin{figure}[t]
\fboxsep0mm
\setlength{\unitlength}{1cm}
\begin{picture}(8.6,6.8)
\put(0,0){\includegraphics[width=8.6cm]{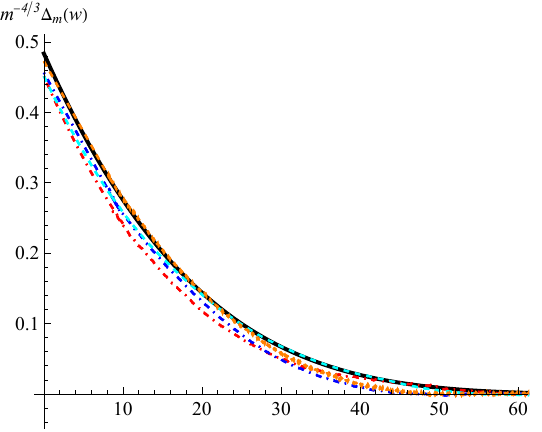}}
\put(3.0,2.5){\includegraphics[width=5.6cm]{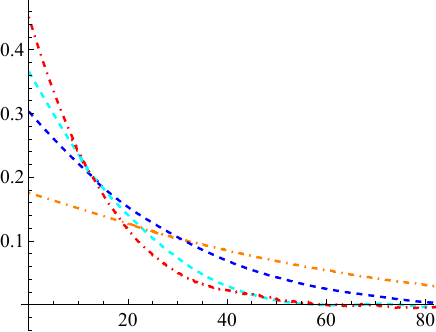}}
\put(7.9,0.8){$w m^{\frac 4 3}$}
\put(7.8,2.3){$w$[nm]}
\put(2.35,6.5){$\Delta_m(w)[\rm pN^2]$}
\put(1,1){\color{black}prediction}
\put(1.8,1.3){\color{black}\vector(1,1){1.15}}
\put(5.4,6){\color{red}region 1}
\put(5.2,6){\color{red}\vector(-4,-1){1.5}}
\put(5.4,5.5){\color{cyan}region 2}
\put(5.3,5.5){\color{cyan}\vector(-4,-1){1.5}}
\put(5.4,5){\color{blue}region 3}
\put(5.3,5){\color{blue}\vector(-4,-1){1.3}}
\put(5.4,4.5){\color{orange}region 4}
\put(5.3,4.5){\color{orange}\vector(-2,-1){1.0}}
\end{picture}
\caption{Inset: The function $\Delta_m(w)$ for the  four regions   changes with the measured  $m$   (see Fig~\ref{f:Fig5}).
Main: Collapse of $\Delta_m(w)$  according to \Eq{DeltaMDef} with $\zeta=4/3$. In black we show the theoretical $\Delta_m(w)$, with $\rho_m = 29(3) \rm nm$ as predicted by the microscopic disorder.} 
\label{f:Fig4}
\end{figure}

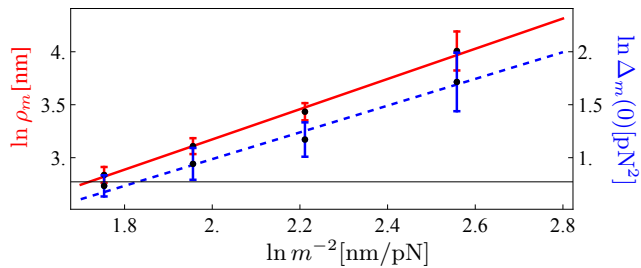
\begin{figure}[t]
\fboxsep0mm
\setlength{\unitlength}{1cm}
{\begin{picture}(8.6,6.4)
\put(0,3.3){\includegraphics[width=8.6cm]{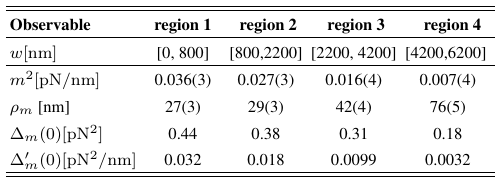}}
\put(0.5,0.3){\includegraphics[width=7.5cm]{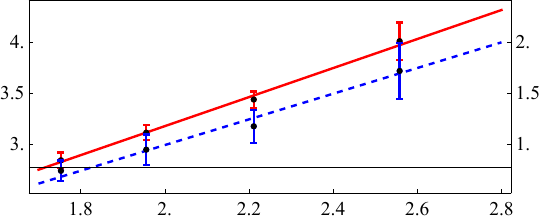}}
\put(0.1,1.5){\rotatebox{90}{$\red \ln \rho_m$\red [nm]}}
\put(8.1,3.0){\rotatebox{-90}{\blue $\ln\Delta_m(0)[\rm pN^2]$}}
\put(3.5,0.){$\ln  \textstyle  {m^{-2}} [\rm nm/pN]$}
\end{picture}}
\caption{Top:  Properties of the force correlator for  the four regions  in Fig.\ \ref{f:Fig2}. The  correlation length $\rho_m=C \Delta_m(0)/\Delta^\prime_m(0)$, with  $C=\tilde{\Delta}_m(0)/\tilde{\Delta}^\prime_m(0)=1.36$, see \Eq{DeltaSinai}. Bottom: The scaling with $m$ of $\rho_m $ (red, solid), $\zeta = 1.41 \pm  0.10$,  and $\Delta_m(0)$ (blue dashed),  $\zeta = 1.29 \pm 0.08$. Their mean  $\zeta=1.34\pm0.06$  agrees with the expected value, $\zeta=4/3$.} 
\label{f:Fig5}
\end{figure}

We   now check two predictions of the theory: the result \eq{DeltaSinai} and the FRG scaling relation \eq{DeltaMDef}. In particular, the scaling function $\tilde \Delta$ only depends on the dimensionless combination $w/\rho_m\sim w m^{\zeta}$, and its amplitude  is universal.   The inset of Fig.~\ref{f:Fig4} shows $\Delta_m(w)$ for the four regions where $\rho_m$ increases while the molecule is unzipped and $m^2$ decreases. In Fig. \ref{f:Fig4} we test the scaling law \eq{DeltaMDef} with $\zeta=4/3$, as predicted for Sinai's model. 
We can also determine the value of $\zeta$ independently of the collapse in Fig.~\ref{f:Fig4}. In  Fig.~\ref{f:Fig5} 
we show results for the scaling of the correlation length $\rho_m$  and amplitude $\Delta_m(0)$ with $m$. We get $\zeta = 1.41 \pm  0.10$ and $\zeta = 1.29 \pm 0.08$ from the scaling of $\rho_m$ and  $\Delta_m(0)$, respectively, giving an average of $\zeta=1.34\pm0.06$ in agreement with the expected value $\zeta=4/3$.
Details are given in 
Supp.~Fig.~\ref{Fig8}.

We can go one step further: In random-field systems, the correlations of the potential $V(u)$ grow linearly at large $u$-distances, $\frac{1}{2} \overline{[V(u)-V(u^\prime)]^2} \simeq \sigma |u-u^\prime|$. The constant $\sigma$ is related to the force correlator $\Delta_m$ by
\be \label{consv}
\sigma= \int_{0}^\infty \Delta_{\infty}(u) \rmd u \equiv \int_{0}^\infty \Delta_m(w) \rmd w\,\,.
\ee
This relation holds for the microscopic $\Delta_{\infty}(u)$ and the measured $\Delta_m(w)$, as the area under $\Delta_m(w)$ is preserved by the RG flow, as previously discussed. A constant $\sigma$ in \Eq{consv} implies $\zeta=4/3$ for all $m$ in \Eq{DeltaMDef}. \Eq{consv} then yields the analytic prediction 
\begin{equation} \label{CorrPred}
\rho_m =  \left[ \frac{ \int_{w>0}  \Delta_{\infty}(w) }{ m^{4} \int_{\sf w>0} \tilde \Delta(\sf w) }\right]^{1/3} .
\end{equation}
In Supp.~Mat.~Sec.~\ref{s:Distribution of binding energies}, we discuss how the microscopic correlator $\Delta_{\infty}(w)$ can be obtained from the binding energies, using our estimate of  $\Delta_{\infty}(0) \approx 10(2) {\rm pN} ^2$, which decays to half this value for BP distance 1, and to 0 for BP distance 2, corresponding to $1.6 {\rm nm}$. A linear interpolation of $\Delta_{\infty}(u)$ between these values gives $\sigma =   8(2) {\rm pN}^2 \cdot \rm {nm}$    in \Eq{consv}. Using $\int_{\sf w>0} \tilde \Delta({\sf w})=0.252$ from \Eq{DeltaSinai}, and substituting in \Eq{CorrPred} gives $ \rho_m= 29(3) {\rm nm}$  for region 1 in agreement with the value previously obtained ($\rho_m \approx 27$nm for $m^2=0.036$ pN/nm in Fig.~\ref{f:Fig5}). In  Fig.~\ref{f:Fig4} (main) we show the predicted force correlator (black curve) with the predicted $ \rho_m= 29(3) {\rm nm}$.

\medskip
\noindent{\it Conclusions.}
We   tested Sinai's model   of equilibrium force correlations and their universality in DNA unzipping experiments. In DNA the binding energies between   base pairs are correlated up to two base pairs, making it a suitable realization of Sinai's model. We   experimentally measured the roughness exponent $\zeta$ finding agreement with Sinai's prediction, $\zeta=4/3$.
While predictions for critical exponents are commonplace, far more difficult is to predict the amplitude and the correlation length of correlation functions in critical phenomena. Here we show that the amplitude of force correlations and its correlation length can be predicted from the effective stiffness of the molecular construct $m^2$ and the energy parameters of the nearest-neighbour model in DNA thermodynamics  \cite{Zuker2003,LorenzBernhartHoner-zu-SiederdissenTaferFlammStadlerHofacker2011}. We get experimental values for $\rho_m$ that are within 10$\%$  of the predicted ones: e.g.~for region 1, $\rho_m\approx 27{\rm nm}$ (measured) versus $\rho_m\approx 29{\rm nm}$  (predicted). 

It is interesting to compare our unzipping experiment to the peeling of complementary RNA-DNA strands \cite{WieseBercyMelkonyanBizebard2019}. Peeling is a highly irreversible process belonging to the depinning universality class. It is characterized by a significantly larger effective stiffness $m^2$, and a larger correlation length of about 186 BP as compared to the 26 to 77 BP of DNA unzipping. 
The high energies required for DNA peeling make the $T=0$ nonequilibrium depinning transition relevant there, whereas for DNA unzipping thermal fluctuations occur in equilibrium.

Our study can be extended to DNA with chemically modified bases and RNA  \cite{rissone2022stem}.  It would also be  interesting to study DNA sequences  with long-range correlations \cite{slutsky2004diffusion} and with periodically repeated motifs, a physical realization of periodic disorder relevant  for charge-density waves. Finally, one could  consider dynamical effects, e.g. upon temperature changes \cite{sales2003temperature} using a temperature-jump optical trap \cite{de2015temperature}.  Overall, single-molecule unzipping offers exciting possibilities to experimentally investigate critical phenomena in random polymers. 



\medskip

\noindent{\it Acknowledgements.}
C. ter Burg and K.~Wiese were supported by LabEx ENS-ICFP.  P.R. was supported by the Angelo della Riccia foundation.  M.R.-P. and F. R.  were supported by Spanish Research Council Grant PID2019-111148GB-I00 and F. R. by the Institució Catalana de Recerca i Estudis Avançats, Academia Prize 2018.

\noindent{\it Author contributions.}  C.T.B and K.J.W developed the field theory. F.R. derived the equation of motion from the experimental model. C.T.B analysed the data. P.R. obtained the experimental data and M.R. set the optical tweezer instrument.  All authors contributed to the writing of  the paper.

\ifx\doi\undefined
\providecommand{\doi}[2]{\href{http://dx.doi.org/#1}{#2}}
\else
\renewcommand{\doi}[2]{\href{http://dx.doi.org/#1}{#2}}
\fi
\providecommand{\link}[2]{\href{#1}{#2}}
\providecommand{\arxiv}[1]{\href{http://arxiv.org/abs/#1}{#1}}
\providecommand{\hal}[1]{\href{https://hal.archives-ouvertes.fr/hal-#1}{hal-#1}}
\providecommand{\mrnumber}[1]{\href{https://mathscinet.ams.org/mathscinet/search/publdoc.html?pg1=MR&s1=#1&loc=fromreflist}{MR#1}}


%

\appendix
%

\begin{widetext}

\newpage

\centerline{\bf \large Experimental test of Sinai's model  in DNA unzipping}

\medskip

\centerline{\bf \large Supplementary Material}

\medskip

\centerline{Cathelijne ter Burg${^{1}}$,  Paolo  Rissone${^{2}}$, Marc Rico-Pasto${^2}$, Felix Ritort${^{2,3}}$ and Kay J\"org Wiese${^1}$}
\begin{center}
\vspace{-5mm}

{\it\small \mbox{{${}^1$}Laboratoire de Physique de l'E\'cole Normale Sup\'erieure, ENS, Universit\'e PSL, CNRS, Sorbonne Universit\'e,} \mbox{Universit\'e Paris-Diderot, Sorbonne Paris Cit\'e, 24 rue Lhomond, 75005 Paris, France.}
  \mbox{{${}^2$}Small Biosystems Lab, Condensed Matter Physics Department, Universitat de Barcelona,} \mbox{Carrer de Mart\'i i Franqu\`{e}s 1, 08028 Barcelona, Spain.}
  \mbox{{${}^3$}Institut de Nanoci\`encia i Nanotecnologia (IN2UB), Universitat de Barcelona, 08028 Barcelona, Spain}}
\end{center}

\renewcommand\appendixname{Section}
\section{Experimental details}
\label{SI:Experiment}
We use laser optical-tweezers (LOT) to unzip a 6.8k base-pair (BPs) DNA hairpin consisting of a stem of fully complementary Watson-Crick base-pairs ending with a tetra-loop ACTA. The hairpin strands are terminated with short (29 BP) double-stranded DNA (dsDNA) handles, one labeled with a digoxigenin tail (DIG) and one with a biotin tail (BIO). In our setup the DIG-labeled and BIO-labeled handles are tethered to anti-DIG (AD) and streptavidin-coated (SA) beads, respectively. The AD bead is optically trapped while the SA one is hold by air suction at the tip of a glass micro-pipette. In a typical unzipping experiment, the optical trap is repeatedly moved back and forth relative the fixed micro-pipette at a constant loading rate. At the initial trap position, the molecule starts in its folded dsDNA configuration (Fig.\ \ref{fig:ExpSetup}(a)). As the optical trap is moved away from the pipette the unzipping reaction proceeds and the molecule gradually unzips (Fig.\ \ref{fig:ExpSetup}(a)) until it is completely unfolded and the two single-strands (ssDNA) are fully stretched (Fig.\ \ref{fig:ExpSetup}(c)). At this point the reverse reaction starts (rezipping) and the molecule steadily refolds into its dsDNA form (dashed arrows). The force versus trap position $F(w)$ sampled during this process shows a saw-tooth pattern which depends on the DNA hairpin sequence. The fact that the unzipping and rezipping saw-tooth force-distance curves coincide (Fig.\ \ref{fig:ExpSetup} (b)) means that the processes are in equilibrium. Here, we unzipped/rezipped the 6.8kpbs DNA hairpin at room temperature (298K) in a wide range of monovalent salt conditions (10mM, 25mM, 50mM, 100mM, 250mM, 500mM and 1000mM NaCl). Fig.\ \ref{fig:ExpSetup} (c) shows unzipping curves for a range of different salt concentrations.

\begin{figure}[h]
\includegraphics[width=0.7\textwidth]{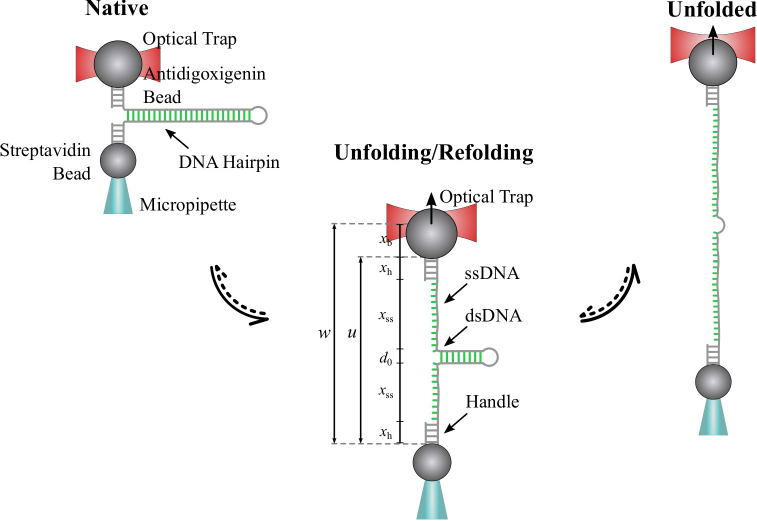}
\caption{Schematic description of the unzipping experiment as reported Supp.~Mat.~Sec.~\ref{SI:Experiment} and representation of the parameters introduced in the computation in Supp.~Mat.~Sec.~\ref{SupplEffTheory}. As reported in the text, $x_{\rm h}$ is the extension of the handles, $x_{\rm b}$ is the displacement of the bead from the center of the optical trap, $d_0$ is the diameter of the folded double-helix ($\sim 2$nm) and $x_{\rm ss}$ is the extension of the ssDNA.}
\label{fig:ExpSetup} 
\end{figure}

\section{Model for the unzipping experiment}
\label{SupplEffTheory}

\noindent Given the   setup in Fig.~\ref{fig:ExpSetup} (middle panel), at a fixed trap-pipette distance $w$ one can write
\begin{equation}
\label{eq:w}
w = x_{\rm b} + u , 
\end{equation}
where $x_{\rm b}$ is the distance of the bead from the center of the optical trap and $u$ is the total length of the molecular construct. The latter can be written as
\begin{equation}
\label{eq:u}
u = 2x_{\rm h}  + d_0 + 2x_{\rm ss}(n), 
\end{equation}
where $x_{\rm h}$ is the extension of the dsDNA handle, $d_0$ is the diameter of the hybridized hairpin and $x_{\rm ss}(n)$ is the extension of the ssDNA. The molecular construct has two short handles, each of 29BP $\equiv$ 10nm in length, much shorter than the persistence length of dsDNA (50 nm) and are very rigid at the unzipping force $(\approx 15{\rm pN})$. Moreover, their overall extension is negligible compared to $x_{\rm ss}(n)$. Also we can neglect $d_0 \approx 2{\rm nm}$. The total length of the molecular construct can be approximated as
\begin{equation}
\label{eq:u_approx}
u \simeq 2x_{\rm ss}(n) \equiv 2nz, 
\end{equation}
with $2n$ the total number of unzipped bases and
\be
z= z(u,n)= \frac u {2n}, 
\ee
 the monomer extension. The total energy is given by 
\begin{align}
\label{eq:Hamiltonian}
\mathcal{H}_{w}(u,n)  \  = &  \frac{1}{2} k_{\rm b} (w - u)^2 + 2n\,\hat U\! \left(\frac{u}{2n} \right) + G_{N-n} . 
\end{align}
The first term is the harmonic potential of the optical trap with stiffness $k_{\rm b}$. The second term, $U_{\rm ss}(u,2n) = 2n \hat U(z)$ is the elastic energy of the ssDNA that behaves as an ideal elastic model. $G_{N-n}$ is the binding energy of the $N-n$ base pairs in the stem.

Given \eqref{eq:Hamiltonian} and  $w$ fixed, $u$ and $n$ are the only two degrees of freedom of the system. Their respective equations of motion are 
\begin{subequations}
\begin{align}
\label{eq:EqsMotion_n}
\gamma_{u} \frac{\partial u}{\partial t} = - \frac{\partial \mathcal{H}_{w}}{\partial u} + \eta_{u},  \\
\label{eq:EqsMotion_u}
\gamma_{n} \frac{\partial n}{\partial t} = - \frac{\partial \mathcal{H}_{w}}{\partial n} + \eta_{n}, 
\end{align}
\end{subequations}
where $\gamma_{u}$ and $\gamma_{n}$ are constants (viscosities), and $\eta_{u}$ and $\eta_{n}$ are $\delta$-correlated Brownian noises with correlations  $\langle \eta_{u}(t) \eta_{u}(t^{\prime}) \rangle = \langle \eta_{n}(t) \eta_{n}(t^{\prime}) \rangle= 2k_{\rm B}T \delta(t - t^{\prime})$.   Mechanical equilibrium for the bead implies 
\begin{equation}
\label{eq:MechEquilibrium}
-\frac{\partial \mathcal{H}_{w}}{\partial u} = 0 \Rightarrow k_{\rm b}(w-u) = f_{1}\left(\frac{u}{2n} \right) , 
\end{equation}
where $f_{1}(z) = -\hat U^{\prime}(z)  $ is the force acting on the ssDNA.
Note that the timescale for $u$ is much faster than the time scale for $n$, so we can take $u$ in equilibrium and \Eq{eq:MechEquilibrium} holds.  This allows us to set   $\eta_{n}\to 0$.

For a given $w$ and $n$, there is an equilibrium position $u^{*}(w,n)$ for which the force exerted by the optical trap on the bead equals the force exerted by the molecular construct on the bead. For the basepair dynamics we find

 \bea
\label{eq:bpDynamics1}
- \frac{\partial \mathcal{H}_{w}}{\partial n} &= & - 2\hat U\!\left(\frac{u}{2n}\right) + \frac{u}{n} \hat U^{\prime} \!\left(\frac{u}{2n}\right) +  G^{\prime}_{N-n}  \nn \\ 
&  = &  -2\hat U(z) + 2z \hat U^{\prime} (z)  + G^{\prime}_{N-n}  \nn \\ 
&= &\text{ }  2\int_{0}^{z}f_{1}({  z'})dz^{\prime} -2z {  f}_{1}(z)   +  G^{\prime}_{N-n}  \nn \\ 
& = & - 2\int_{0}^{f_{1}(z)}z(f^{\prime})\rmd f^{\prime}  +  G^{\prime}_{N-n} \nn  \\ 
& = & I_{w}(f(n)) + G^{\prime}_{N-n},  
\eea
where 
\be
I_w(f) := \text{ } -2\int_{0}^{f}z(f^{\prime})\rmd f^{\prime} \label{Iwfdef}.
\ee
For a given $w$ and $u^{*}(w, n)$, there is an equilibrium $n^{*}$,  so that  \Eq{eq:bpDynamics1} gives 
\begin{equation}
\label{eq:Eq_n}
- \frac{\partial \mathcal{H}_w(n)}{\partial n}  \bigg |_{n^{*}}  =   I_w(n^{*}) + G^\prime (N - n^{*}) = 0 \, \Rightarrow \, I_w(n^{*}) =- G^\prime (N - n^{*}), 
\end{equation}
where $n^{*}$ is the value at which the absolute minimum is attained.  Expanding $I_w(n)$ around $n^*$ and substituting the previous result one gets
\bea
I_w(n) & = &    I _{w}(n^{*}) + \frac{\partial I_w}{\partial n}\biggl|_{n^*} (n - n^*) + \mathcal{O}(n - n^*)^2  \\ 
& = &  -G^\prime (N - n^{*}) + \frac{\partial I_w}{\partial n}\biggl|_{n^*} (n - n^*) + \mathcal{O}(n - n^*)^2. 
\eea
Let us focus on the first-order term. Using \Eq{Iwfdef}, it can be rewritten as 
\begin{equation}
\label{eq:IntegralApprox1}
\frac{\partial I_w}{\partial n} =  \frac{\partial I_w(n)}{\partial f_{1}}\frac{\partial f_{1}}{\partial n} = - 2z_1 \frac{\partial f_{1}}{\partial n}, 
\end{equation}
where $z_1 := z(f_{1})$. 
Taking a derivative w.r.t. $n$ of  \Eq{eq:MechEquilibrium} we obtain
\begin{equation}
\label{eq:IntegralApprox1.1}
- k_{\rm b} \frac{\partial u}{\partial n }  =  \frac{\partial f_1\left(\frac{u}{2n} \right)}{\partial n}  =  k_1 \!\left( \frac{u}{2n} \right)  \left[ \frac{1}{2n}\frac{\partial u}{\partial n} - \frac{u}{2n^2} \right], 
\end{equation}
where
\be
k_1 \equiv k_1(z):=  f_1'(z).
\ee  
Solving for $\frac{\partial u}{\partial n }$ yields
\begin{equation}
\label{eq:IntegralApprox1.2} 
\frac{\partial u}{\partial n }  
= \frac{k_{1}u}{ n  \left(2n k_{\rm b}+ {k_{1}} \right)}. 
\end{equation}
Using \Eq{eq:IntegralApprox1}, and the first equality of \Eq{eq:IntegralApprox1.1} gives 
\be
\frac{\partial I_w}{\partial n}  =- 2z_1 \frac{\partial f_{1}}{\partial n} = - 2z_1 k_{\rm b}\frac{\partial u}{\partial n} = 
 - \frac{  2z_1 k_{\rm b} k_{1}u}{ { n  \left(2n k_{\rm b}+ {k_{1}} \right)} } =   - \frac{4 z_1^2 k_{\rm b} k_1}{k_1 + 2n k_{\rm b}}.
\ee
The equation of motion for the basepairs dynamics in \Eq{eq:EqsMotion_u} can therefore we written as
\bea
\label{eq:bpDynamics3}
\gamma_n \frac{\partial n}{\partial t} &=&I_w(n) +G'(N-n)+ \eta_n\simeq I_w(n^*)  + \frac{\partial I_w(n)}{\partial n}\Big|_{n=n^*}  (n-n^*) +G'(N-n) + \eta_n\nn\\
&=&  \frac{\partial I_w}{\partial n} \Big|_{n=n^*}  (n-n^*) - G'(N-n^*) +G'(N-n) + \eta_n
\nn\\
&=& - \frac{4z_1^2 k_{\rm b} k_1}{k_1 + 2n k_{\rm b}} (n - n^*) - G^\prime(N - n^*) + G^\prime(N - n) + \eta_n  .   
\eea
Define
\be
G(N-n) = \sum_{i=N-n}^N g_i. 
\ee
The random forces are
\be
f_n = -G'(N-n) = -g_{N - n} \equiv - g(N - n). 
\ee 
We observe that 
\begin{equation}
u = 2 \langle z_1 \rangle  n,  \qquad  u^* = 2  \langle z_1 \rangle  n^*, \qquad \frac{\partial n}{\partial t } = \frac{1}{2\langle z_1 \rangle  } \frac{\partial u}{\partial t}, 
\end{equation}
where $\langle z_1 \rangle$ denotes a statistical average over $z_1$.

The final equation of motion for $u $ can be rewritten as 
\bea
\tilde{\gamma}_n \frac{\partial u}{\partial t}& = &  m^2 (  u^* - u) + F(u ) - F(u^*) \\ 
&\equiv  &   m^2 (  w - u) + F(u ),   
\eea
where 
\begin{align}
w \ = & \text{ } u^* - \frac{1}{m^2} F(u^*),  \\ 
 {\tilde \gamma_n} \  = &  \frac{\gamma_n}{4\langle z_1 \rangle^2} ,   \\ 
  F(u ) \ = &-2\langle z_1 \rangle \, g\!\left(N-\frac {  u}{2\langle z_1 \rangle}\right)  ,   \\ 
  \frac 1 {m^2} \ = &    {\frac 1 { k_{\rm b}  } + \frac {2n} { k_{1} }}   \approx {\frac 1 { k_{\rm b}  } + \frac { u} {\langle z_1 \rangle   k_{1} }} \approx {\frac 1 { k_{\rm b}  } + \frac {w} {\langle z_1 \rangle  k_{1} }} . \label{eq:eff_stiffness2}
\end{align} 
Note that $u^*$ in this equation is simply {\em a minimum}, not necessarily the previously assumed global minimum.  Rescaling $t$, we set $\tilde{\gamma}_n \to 1$. Noticing that $u$ and $ w$ in the experiment are defined up to an overall shift gives equation \eqref{eq:EqMotion} in the main text. 

\section{Distribution of binding energies}
\label{s:Distribution of binding energies}
In Ref.~\cite{JosepHuguetBizarroFornsSmithBustamanteRitort2010,HuguetRibezziCrivellariBizarroRitort2017} the base-pair free energies in DNA where experimentally determined for the 6kBP sequence studied in this paper. Fig. \ref{f:suppsalt}(a)  shows that the distribution of basepair energies is roughly Gaussian, here given for the 1M salt concentration.  
{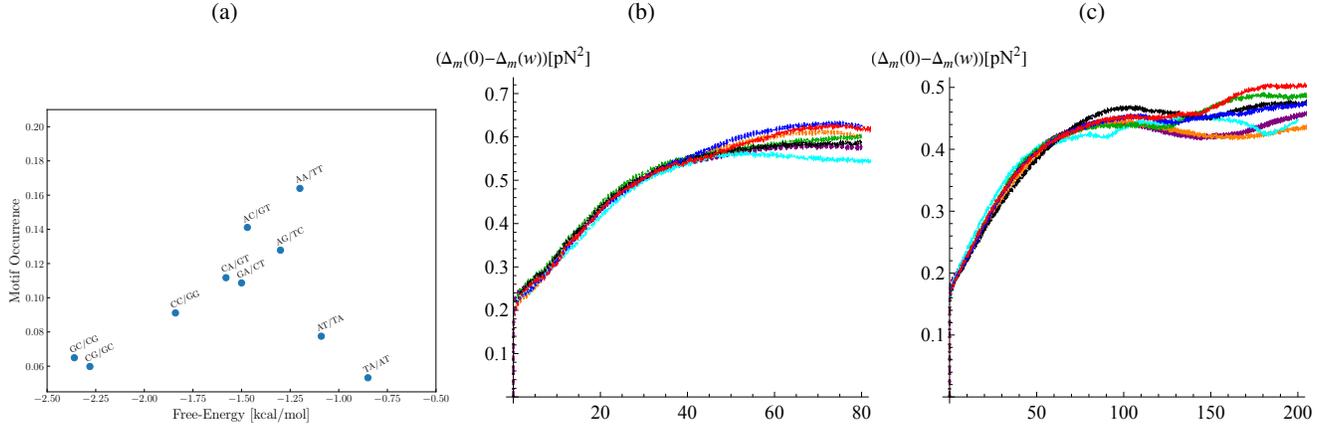
\begin{figure*}[t]
{\setlength{\unitlength}{1cm}\begin{picture}(17.8,5.9)
\put(0,0){\includegraphics[width=.33\textwidth]{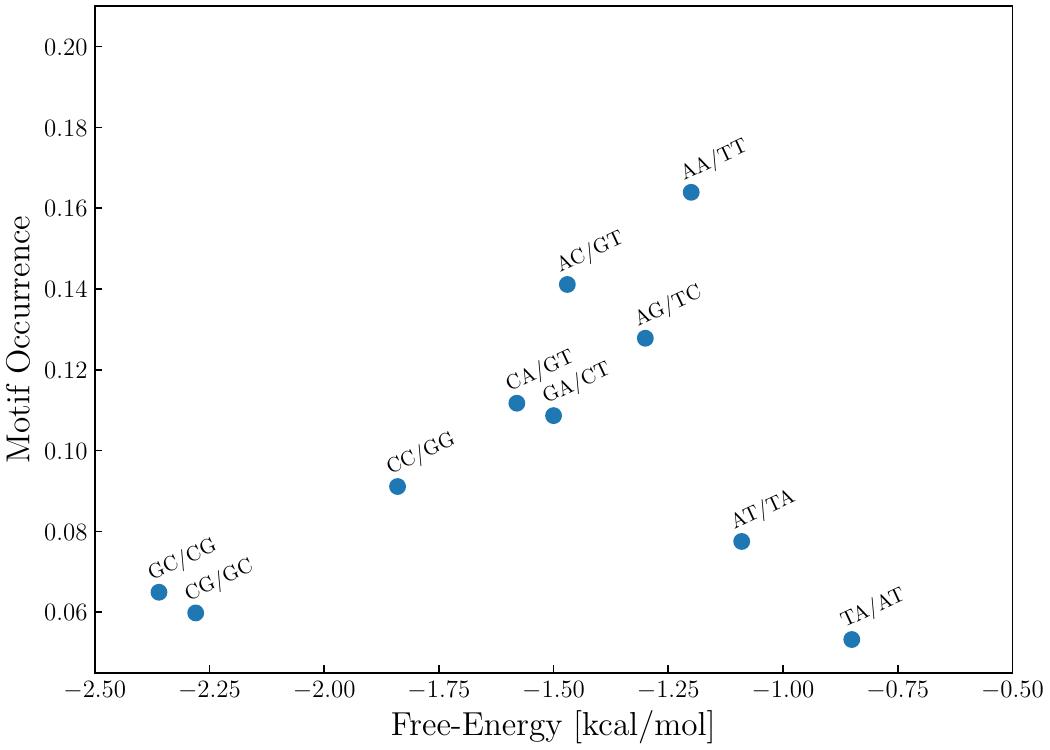}}
\put(5.7,0){\includegraphics[width=.33\textwidth]{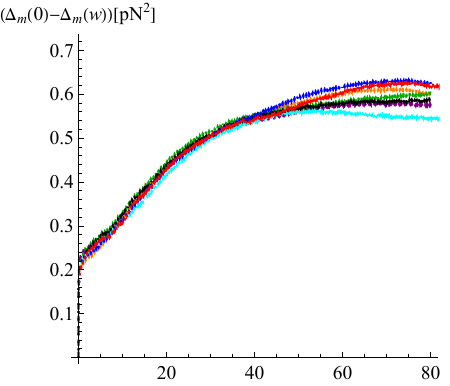}}
\put(11.5,0){\includegraphics[width=.33\textwidth]{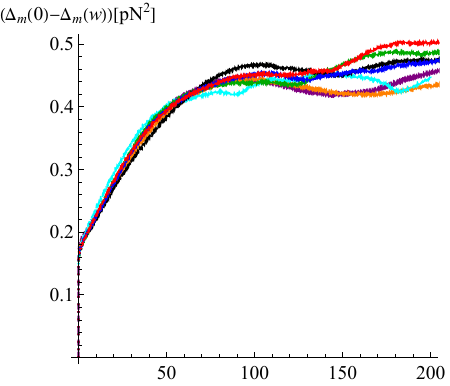}}
\put(2.72,5.4){(a)}
\put(8.25,5.4){(b)}
\put(14.25,5.4){(c)}
\end{picture}}
\caption{(a) Distribution of the nearest-neighbour binding energies at 1M NaCl salt concentration for the 6 kBP sequence studied here. For the first (b) and third (c) region, the averaged $\Delta_{m}$ for each salt concentration. Despite the dependence of the pinning force on the salt concentration, the correlation length is not visibly affected by the salt concentration.  Color code as in Fig. \ref{f:Fig1} (c). }
\label{f:suppsalt}
\end{figure*}}%
Using these binding energies  one can  evaluate the microscopic disorder $\Delta_\infty(w)$. One uses that $1 {\rm kcal/mol} = 6.944 {\rm pN \cdot nm} $ and converts to forces by dividing by the base-pair length of  $\approx 0.8 {\rm nm}$ \cite{AlemanyRitort2014}. The resulting $\Delta_\infty(w)$ decays to zero on a scale of $2 {\rm BP} \approx 1.6\, \rm nm$. \\
\\
\noindent The   FRG implies that  $\int_{w>0} \Delta(w) $ is independent of $m^2$. As a consequence
\bea
&& \int_0^\infty \rmd w \, \Delta_{\infty}(w)  \equiv  \int_0^\infty \rmd w \, \Delta_{m}(w) \nn \\
& &= \int_0^\infty \rmd w\, m^4 \rho_m^2 \tilde \Delta_m(w/\rho_m) = m^4 \rho_m^3 \int_0^\infty \rmd w\, \tilde \Delta_m(w) =  0.252\,   m^4 \rho_m^3 .
 \label{IntCond}
\eea
Solving for $\rho_m$ yields    \Eq{CorrPred} presented in the main text, 
\be
\rho_m =  \left[ \frac{  \int_{w>0}  \Delta_{\infty}(w) }{ m^{4} \int_{\sf w>0} \tilde \Delta(\sf w) }\right]^{1/3} \simeq   \left[ \frac {3.97}  { m^{4}} \int_{w>0}  \Delta_{\infty}(w)   \right]^{1/3} .  \label{rhomPred}
\ee
The equality \eqref{IntCond} is satisfied in our experiment, a strong test of universality. We find  $\int_{w>0} \Delta_m (w) \approx 7.2 (4) {\rm pN}^2 \cdot {\rm nm}$, for the first region,$\int_{w>0} \Delta_m (w) \approx  7.0(4) {\rm pN}^2 \cdot {\rm nm}$ for the second, $\int_{w>0} \Delta_m (w) \approx 7.0 (5) {\rm pN}^2 \cdot {\rm nm} $ for the third and $ \int_{w>0} \Delta_m (w) \approx 7.2 (7) {\rm pN}^2 \cdot {\rm nm}$  for the last region. Our estimate of the microscopic $\int_{w>0} \Delta_\infty(w)$   comes close to this. If we use the binding energies of \cite{HuguetRibezziCrivellariBizarroRitort2017} for the 1{\rm M} salt concentration, we find $\int_{w>0} \Delta_\infty(w) \approx 8 (2)  {\rm pN}^2 \cdot {\rm nm} $\cite{JosepHuguetBizarroFornsSmithBustamanteRitort2010,HuguetRibezziCrivellariBizarroRitort2017}.  For the first region this corresponds to $\rho_m = 29(3) {\rm nm}$ close to the experimentally measured value of 27(3)nm. 

\subsection*{Salt dependence}
The base-pair energies change with  the salt concentration according to 
\begin{equation}
{\rm dG}({\rm[salt]}) = {\rm dG}_0([1{\rm M}]) - m_i \ln({\rm[salt]}), 
\end{equation}
where $\rm{dG}_0$ is the binding energy at a 1M salt concentration. The values for $m_i$ are  given in Table 1 of Ref.~\cite{JosepHuguetBizarroFornsSmithBustamanteRitort2010}. 
Values for $\int_{w>0} \Delta_\infty(w)$ for different salt concentrations are given in Table \ref{tabmic}. One can do a similar analysis, taking into account the proportion of samples per salt concentration. One finds a  somewhat larger value  $\int_{w>0} \Delta_\infty(w) \approx 9 (2)  {\rm pN}^2 \cdot {\rm nm} $ and  $\rho_m = 30(3) {\rm nm}$, still in agreement within error bars.

\begin{table}[!h]
\begin{tabular}{lcccc}
\hline\hline
 {\bf \footnotesize salt concentration [mM]} & ~ {\bf \footnotesize $ \int_{w >0} \Delta_\infty(w)$ } ~ & ~ {\bf \footnotesize  \# samples }~ &~ {\bf \footnotesize }~ &~ {\bf \footnotesize   }~\\
\hline
\footnotesize $1000$ &\footnotesize  8(2)  & \footnotesize 6  &\footnotesize    &\footnotesize   \\

\footnotesize $500$ &\footnotesize 8(2) & \footnotesize  7&\footnotesize   &\footnotesize   \\

\footnotesize $250$  & \footnotesize 8(2)   & \footnotesize 5  &\footnotesize   &\footnotesize  \\

\footnotesize $100$  & \footnotesize 9(2)  & \footnotesize 5 &\footnotesize   &\footnotesize  \\

\footnotesize $50$  & \footnotesize 9(2)  & \footnotesize 4 &\footnotesize   &\footnotesize  \\

\footnotesize $25$  & \footnotesize 10(3) & \footnotesize 3 &\footnotesize   &\footnotesize  \\
\footnotesize $10$  & \footnotesize 10(3)  & \footnotesize  3 &\footnotesize   &\footnotesize  \\

\hline \hline
\end{tabular}
\caption{Properties of the microscopic disorder using the binding energies of \cite{JosepHuguetBizarroFornsSmithBustamanteRitort2010}. In our analysis, we averaged over the different salt concentrations (see main text). The resulting error is small, of the order of 2BP on the total correlation length.}
\label{tabmic}
\end{table}

 \section{RG at finite temperature,   rounding and  deconvolution}
\label{suppUnfolding}

{\begin{figure*}[t]
{\setlength{\unitlength}{1cm}\begin{picture}(17.8,7.2)
\put(0,0){ \includegraphics[width=.49\textwidth]{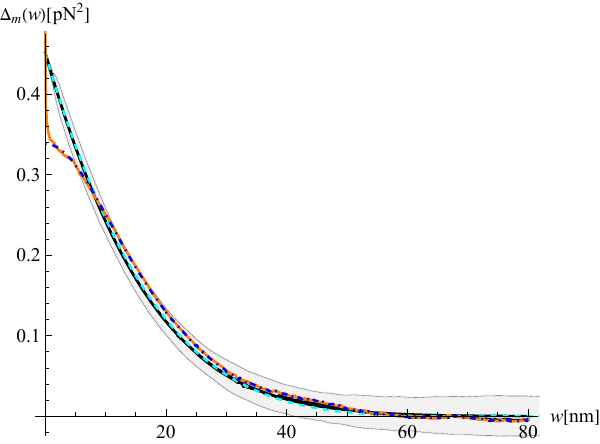}}
\put(8.9,0){ \includegraphics[width=.49\textwidth]{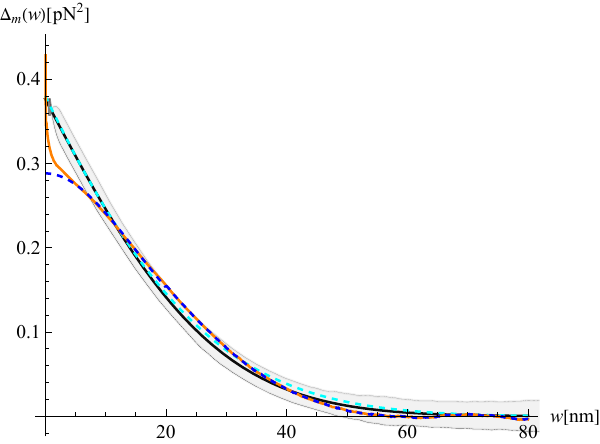}}
\put(4.4,7){(a)}
\put(13.2,7){(b)}

\put(2.5,2.9){\color{orange} {raw data (measurement)}}
\put(2.9,2.9){\color{orange}\vector(-3,-1){0.8}}
\put(0.86, 2.2){\color{blue}\vector(0,1){2.2}}
\put(0.8,1.9){\color{blue} Eq.~\eqref{D3}}
\put(2.1,4.75){\color{black} {extracted $T = 0$ data}}
\put(2.0,4.75){\color{black}\vector(-3,-1){0.85}}
\put(4.8,0.9){\color{cyan} {theory}}
\put(4.7,0.92){\color{cyan}\vector(-3,-1){0.6}}
\put(12.0,2.9){\color{orange} {raw data (measurement)}}
\put(11.9,2.9){\color{orange}\vector(-3,-1){0.8}}
\put(9.86, 2.1){\color{blue}\vector(0,1){1.85}}
\put(9.70,1.8){\color{blue}    Eq.~\eqref{D3}}
\put(11,4.75){\color{black} {extracted $T = 0$ data}}
\put(11.0,4.75){\color{black}\vector(-3,-1){1.0}}
\put(14.3,0.9){\color{cyan} {theory}}
\put(14.2,0.92){\color{cyan}\vector(-3,-1){0.6}}

\end{picture}}
{\setlength{\unitlength}{1cm}\begin{picture}(17.8,7.4)
\put(0,0){ \includegraphics[width=.49\textwidth]{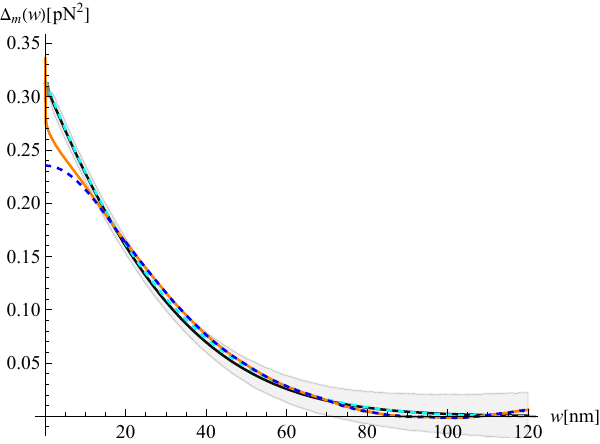}}
\put(8.9,0){ \includegraphics[width=.49\textwidth]{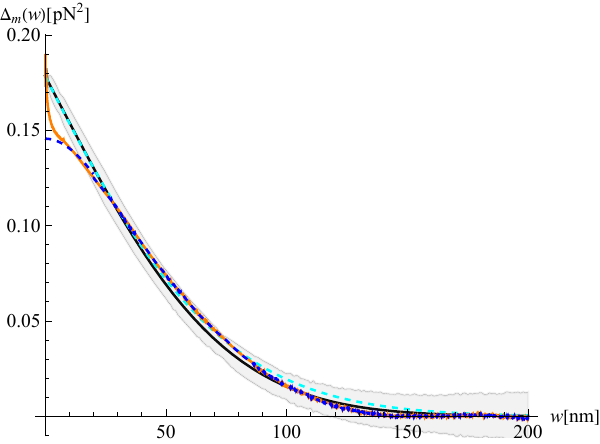}}

\put(2.5,2.9){\color{orange} {raw data (measurement)}}
\put(2.9,2.9){\color{orange}\vector(-3,-1){0.8}}
\put(0.9, 2.2){\color{blue}\vector(0,1){1.85}}
\put(0.8,1.9){\color{blue}    Eq.~\eqref{D3}}
\put(2.1,4.75){\color{black} {extracted $T = 0$ data}}
\put(2.0,4.75){\color{black}\vector(-3,-1){0.85}}
\put(4.8,0.9){\color{cyan} {theory}}
\put(4.7,0.92){\color{cyan}\vector(-3,-1){0.4}}
\put(12.0,2.9){\color{orange} {raw data (measurement)}}
\put(11.9,2.9){\color{orange}\vector(-3,-1){0.7}}
\put(9.86, 2.1){\color{blue}\vector(0,1){2.3}}
\put(9.7,1.8){\color{blue}    Eq.~\eqref{D3}}
\put(11.1,4.75){\color{black} {extracted $T = 0$ data}}
\put(11.0,4.75){\color{black}\vector(-3,-1){0.9}}
\put(14.3,0.9){\color{cyan} {theory}}
\put(14.2,0.92){\color{cyan}\vector(-3,-1){0.55}}

\put(4.4,7){(c)}
\put(13.2,7){(d)}

\end{picture}}
\caption{Comparison of the force correlation $\Delta_m(w)$ for segments 1-4 in (a-d) extracted at $T=0$ (black with grey error bar) with the theoretical prediction (cyan dashed line). The blue line shows the result of \eqref{D3}, testing our convolution and reconstructing the finite-$T$ data using the diffusion kernel which well reproduces the experimental data (orange). One sees that as $m^2$ decreases, the correlation lenght increases and the size of the thermal peak decreases \cite{terBurgWiese2022}.  }
\label{Fig8}
\end{figure*}
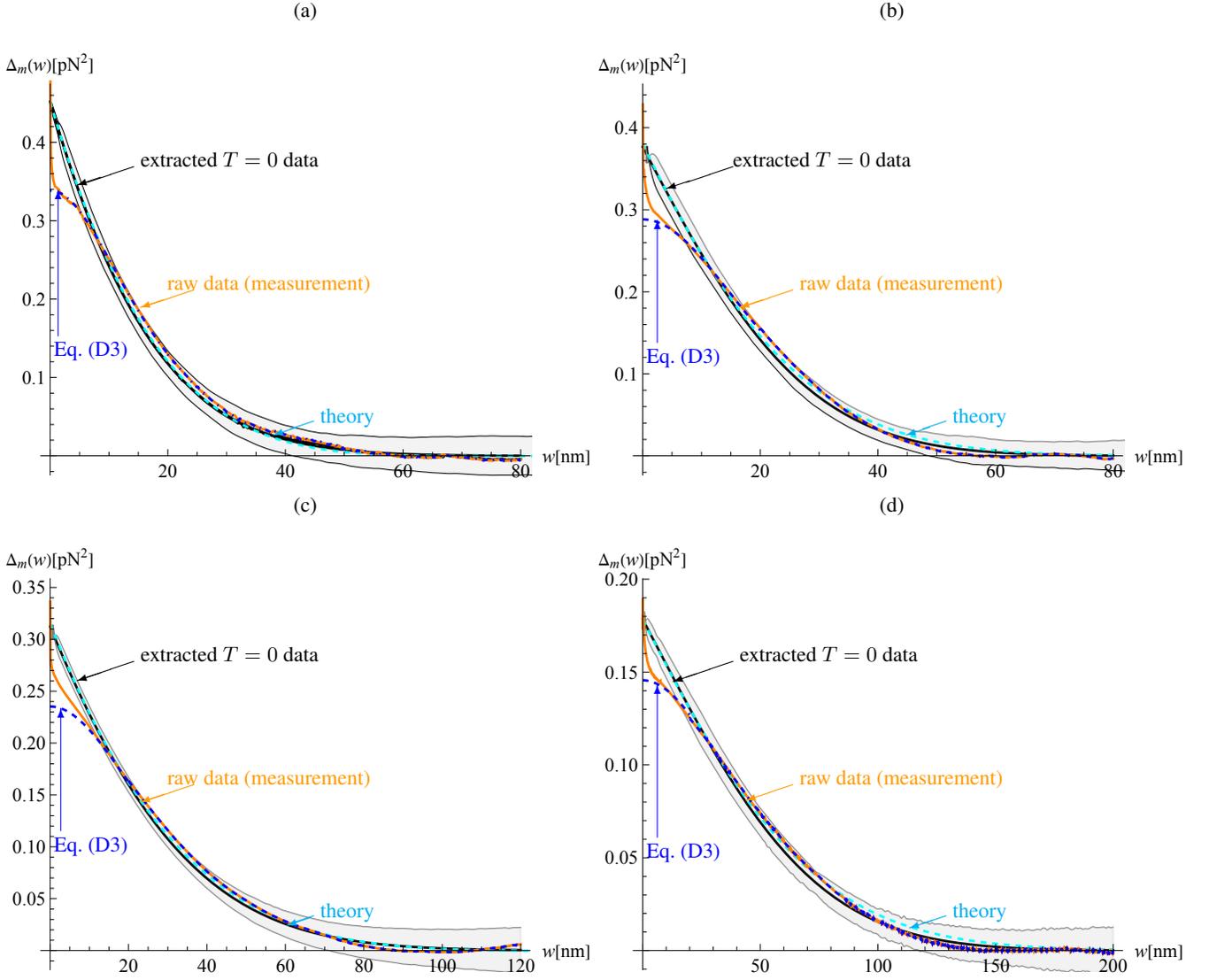}

At finite temperature, the 1-loop FRG equation without rescaling acquires an additional term  
 \be 
-m \partial_m  \Delta_m (w)= 
  -   \frac{1}{2}  \partial_w^2\bigl[ {\Delta}_m(w) {-} {\Delta_m}(0) \bigl] ^2   
+ \tilde{T}_m \Delta_m^{\prime \prime }(w)  \dots  \label{1loopFRGTa}
\ee
In \Eq{1loopFRG} we have written the fixed-point equation for the rescaled version $\tilde \Delta(w) = m^{\epsilon-2 \zeta }\Delta(w m^\zeta)$, 
 \be 
-m \partial_m \tilde \Delta ({\sf w})=    (\epsilon  {-} 2 \zeta) \tilde{\Delta}({\sf w}) + \zeta {\sf w} \tilde{\Delta}^\prime({\sf w})  
  -   \frac{1}{2}  \partial_{\sf w}^2\bigl[ \tilde{\Delta}({\sf w}) {-} \tilde{\Delta}(0) \bigl] ^2   
+ \tilde{T}_m \tilde{\Delta}^{\prime \prime }({\sf w})  \dots  \label{1loopFRGTb}
\ee
What is remarkable about \Eq{1loopFRGTa} is that the RG flow conserves the integral $\int_{w>0} \Delta(w)$, both at vanishing temperature $\tilde T_m=0$ and at  $\tilde T_m>0$. The reason is that the r.h.s.\ of \Eq{1loopFRGTa} is a total derivative. For the random field solution $\zeta=\epsilon/3$ relevant for us, this also holds for the rescaled  \Eq{1loopFRGTb}. 

In \Eq{BdrLayer} we wrote the  finite-temperature solution in the   standard    form
\be\label{D3}
\Delta_{m,T}(w)    \approx    {\cal N}   \Delta_m(\sqrt{w^2 + t^2}), \qquad t  = 
 \frac{6 m^2 k_{\rm B}T  }{\epsilon |\Delta^\prime_m(0)|}.
\ee
As the flow preserves the area, it is important to fix $\ca N$, s.t.\ the integrals on both sides coincide. 
This adds a non-trivial   change in normalization, which cannot be given in closed form. 
Another problem with \Eq{D3} is that given $\Delta_{m,T}(w)$, one can  reconstruct $\Delta_m(w)$ only for $w \ge t$. 
Since   this is a phenomenological approximation, we may propose a different approximation. Namely, to obtain the finite-$t$ solution   by convoluting the zero-temperature solution with an appropriately chosen diffusion kernel,   
%
%
\be\label{DiffKernelsupp} 
\Delta_{m, T}(w)  =    \int_{-\infty}^{\infty} \rmd u\,  \Delta_m(u) R(u-w,\tau), \qquad 
R(u,\tau)  = \frac{1}{ \sqrt{4\pi \tau}} \rme^{-\frac{u^2}{4 \tau }}.  
\ee
A nice property of the convolution prescription in \Eq{DiffKernelsupp} is that   by construction it is area preserving. 
What remains to be done is to fix the   ``diffusion time'' $\tau$. Given the properties of the diffusion kernel, this can analytically be done for 
\be
\Delta_m(w) = \ca C  \rme^{-a w - b w^2}.
\ee
Demanding that $\Delta_{m,T}''(0)/\Delta_{m,T}(0)$ agree yields
\be
\tau = \frac{t^2}{\pi} - \frac{2a(\pi - 2)t^3}{\pi^2} +  \mathcal{O}(t^4).
\ee
The leading-order term only depends on $t$, while the  subleading term contains $a= -\Delta'(0^+)/\Delta(0)\sim 1/\rho_m$. 
In the example of Figs.~\ref{f:Fig3}(a)--(b) we find $\tau= 18.87$ at leading order, and $\tau =22 $ at subleading order. The latter value yields an excellent agreement between approximations \eq{D3} and  \eq{DiffKernelsupp}, and is used in the tests on Fig.~\ref{Fig8}.

\end{widetext}


\begin{thebibliography}{10}

\bibitem{bouchaud1990anomalous}
J.-P. Bouchaud and A. Georges,
\newblock{\em Anomalous diffusion in disordered media: statistical mechanisms, models and physical applications},
\newblock \doi{10.1016/0370-1573(90)90099-N}{\rm Phys. Rep. {\bf
  195} (1990)   127-293}\null.

\bibitem{kardar1998nonequilibrium}
M. Kardar,
\newblock{\em Nonequilibrium dynamics of interfaces and lines},
\newblock \doi{10.1016/S0370-1573(98)00007-6}{\rm Phys. Rep. {\bf
  301} (1998)   85-112}\null.
  

\bibitem{kirkpatrick2015colloquium}
T. R. Kirkpatrick and D. Thirumalai,
\newblock{\em Colloquium: Random first order transition theory concepts in biology and physics},
\newblock \doi{10.1103/RevModPhys.87.183}{\rm Rev. Mod. Phys. {\bf
  87} (2015)   183}\null.
  

\bibitem{schrodinger1944life}
E. Schr\"odinger, 
\newblock{\em What is life?},
\newblock \doi{10.1017/CBO9781139644129}{\rm  Cambridge University Press
  (1944)}.




\bibitem{varn2016did}
D.P. Varn  and J.P. Crutchfield,
\newblock{\em What did Erwin mean? The physics of information from the materials genomics of aperiodic crystals and water to molecular information catalysts and life},
\newblock \doi{10.1098/rsta.2015.0067}{\rm  Philos. Trans. Royal Soc. A {\bf  374-2063}
  (2016)   20150067}\null.



\bibitem{Sinai1983}
Y.G. Sinai,
\newblock {\em The limiting behaviour of a one-dimensional random walk in a
  random environments},
\newblock \doi{10.1137/1127028}{\rm Theory Probab. Appl. {\bf 27} (1983)
  256--268}\null.

\bibitem{LeDoussalWieseMoulinetRolley2009}
P.~Le Doussal, K.J. Wiese, S.~Moulinet  and E.~Rolley,
\newblock {\em Height fluctuations of a contact line: {A} direct measurement of
  the renormalized disorder correlator},
\newblock \doi{10.1209/0295-5075/87/56001}{\rm EPL {\bf 87} (2009)
  56001}\null,
\newblock \arxiv{arXiv:0904.4156}.

\bibitem{terBurgBohnDurinSommerWiese2021}
C.~ter Burg, F.~Bohn, F.~Durin, R.L. Sommer  and {K.J.} Wiese,
\newblock {\em Force correlations in disordered magnets},
\newblock \doi{10.1103/PhysRevLett.129.107205}{\rm Phys. Rev. Lett. {\bf 129}
  (2022)   107205}\null,
\newblock \arxiv{arXiv:2109.01197}.



\bibitem{WieseBercyMelkonyanBizebard2019}
K.J. Wiese, M.~Bercy, L.~Melkonyan  and T.~Bizebard,
\newblock {\em Universal force correlations in an {RNA-DNA} unzipping
  experiment},
\newblock \doi{10.1103/PhysRevResearch.2.043385}{\rm Phys. Rev. Research {\bf
  2} (2020)   043385}\null,
\newblock \arxiv{arXiv:1909.01319}.


\bibitem{supp}
Supplemental Material for this letter, see page 6. 


\bibitem{HuguetFornsRitort2009}
J.~M. Huguet, N.~Forns  and F.~Ritort,
\newblock {\em Statistical properties of metastable intermediates in DNA
  unzipping},
\newblock \doi{10.1103/PhysRevLett.103.248106}{\rm Phys. Rev. Lett. {\bf 103}
  (2009)   248106}\null.

\bibitem{AlemanyRitort2014}
A.~Alemany and F.~Ritort,
\newblock {\em Determination of the elastic properties of short ssDNA molecules
  by mechanically folding and unfolding {DNA} hairpins},
\newblock \doi{10.1002/bip.22533}{\rm Biopolymers {\bf 101}
  (2014)   1193--1199}.


\bibitem{Wiese2021}
K.J. Wiese,
\newblock {\em Theory and experiments for disordered elastic manifolds,
  depinning, avalanches, and sandpiles},
\newblock \doi{10.1088/1361-6633/ac4648}{Rep. Prog. Phys. 85 (2022) 086502 (133pp)},
\newblock \arxiv{arXiv:2102.01215}.

\bibitem{BalentsLeDoussal2004}
L.~Balents and P.~Le Doussal,
\newblock {\em Thermal fluctuations in pinned elastic systems: field theory of
  rare events and droplets},
\newblock \doi{10.1016/j.aop.2004.10.001}{\rm Ann. Phys. (NY) {\bf 315} (2005)
   213--303}\null,
\newblock \arxiv{cond-mat/0408048}.

\bibitem{ChauveGiamarchiLeDoussal2000}
P.~Chauve, T.~Giamarchi  and P.~Le Doussal,
\newblock {\em Creep and depinning in disordered media},
\newblock \doi{10.1103/PhysRevB.62.6241}{\rm Phys. Rev. B {\bf 62} (2000)
  6241--67}\null,
\newblock \arxiv{cond-mat/0002299}.

\bibitem{DSFisher1986}
D.S. Fisher,
\newblock {\em Interface fluctuations in disordered systems: {$5-\epsilon$}
  expansion},
\newblock \doi{10.1103/PhysRevLett.56.1964}{\rm Phys. Rev. Lett. {\bf 56}
  (1986)   1964--97}\null.

\bibitem{NattermannStepanowTangLeschhorn1992}
T.~Nattermann, S.~Stepanow, L.-H. Tang  and H.~Leschhorn,
\newblock {\em Dynamics of interface depinning in a disordered medium},
\newblock \doi{10.1051/jp2:1992214}{\rm J. Phys. II (France) {\bf 2} (1992)
  1483--8}\null.

\bibitem{NarayanDSFisher1993a}
O.~Narayan and D.S. Fisher,
\newblock {\em Threshold critical dynamics of driven interfaces in random
  media},
\newblock \doi{10.1103/PhysRevB.48.7030}{\rm Phys. Rev. B {\bf 48} (1993)
  7030--42}\null.

\bibitem{ChauveLeDoussalWiese2000a}
P.~Chauve, P.~Le Doussal  and K.J. Wiese,
\newblock {\em Renormalization of pinned elastic systems: How does it work
  beyond one loop?},
\newblock \doi{10.1103/PhysRevLett.86.1785}{\rm Phys. Rev. Lett. {\bf 86}
  (2001)   1785--1788}\null,
\newblock \arxiv{cond-mat/0006056}.

\bibitem{LeDoussalWieseChauve2002}
P.~Le Doussal, K.J. Wiese  and P.~Chauve,
\newblock {\em 2-loop functional renormalization group analysis of the
  depinning transition},
\newblock \doi{10.1103/PhysRevB.66.174201}{\rm Phys. Rev. B {\bf 66} (2002)
  174201}\null,
\newblock \arxiv{cond-mat/0205108}.

\bibitem{LeDoussalWieseChauve2003}
P.~Le Doussal, K.J. Wiese  and P.~Chauve,
\newblock {\em Functional renormalization group and the field theory of
  disordered elastic systems},
\newblock \doi{10.1103/PhysRevE.69.026112}{\rm Phys. Rev. E {\bf 69} (2004)
  026112}\null,
\newblock \arxiv{cond-mat/0304614}.

\bibitem{le2006can}
P.~Le Doussal, K.J. Wiese, E. Raphael and R. Golestanian,
\newblock {\em Can nonlinear elasticity explain contact-line roughness at depinning?},
\newblock \doi{10.1103/PhysRevE.69.026112}{\rm Phys. Rev. Lett. {\bf 96} (2006)
  015702}\null,
  

\bibitem{WieseHusemannLeDoussal2018}
K.J. Wiese, C.~Husemann  and P.~{Le Doussal},
\newblock {\em Field theory of disordered elastic interfaces at 3-loop order:
  The $\beta$-function},
\newblock \doi{10.1016/j.nuclphysb.2018.04.013}{\rm Nucl. Phys. B {\bf 932}
  (2018)   540--588}\null,
\newblock \arxiv{arXiv:1801.08483}.

\bibitem{HusemannWiese2017}
C.~Husemann and K.J. Wiese,
\newblock {\em Field theory of disordered elastic interfaces to 3-loop order:
  Results},
\newblock \doi{10.1016/j.nuclphysb.2018.04.015}{\rm Nucl. Phys. B {\bf 932}
  (2018)   589--618}\null,
\newblock \arxiv{arXiv:1707.09802}.



\bibitem{zinati2022stochastic}
R.B.A. Zinati, C. Duclut, S. Mahdisoltani, A. Gambassi and R. Golestanian,
\newblock {\em Stochastic dynamics of chemotactic colonies with logistic growth},
\newblock \doi{10.1209/0295-5075/ac48c9}{\rm  EPL {\bf  136}
  (2022)   50003}\null,


\bibitem{terBurgWiese2022}
C.~ter Burg and {K.J.} Wiese,
\newblock {\em Force-force correlator for driven disordered systems at finite
  temperature},
\newblock (2022),
\newblock \arxiv{arXiv:2201.12652v1}.


\bibitem{LeDoussal2009}
P.~Le Doussal,
\newblock {\em The Sinai model in presence of dilute absorbers},
\newblock \doi{10.1088/1742-5468/2009/07/P07032}{\rm J. Stat. Mech. (2009)
  P07032}\null,
\newblock \arxiv{arXiv:0906.0267}.



\bibitem{Zuker2003}
M.~Zuker,
\newblock {\em Mfold web server for nucleic acid folding and hybridization
  prediction},
\newblock \doi{10.1093/nar/gkg595}{\rm Nucleic Acids Res. {\bf 31} (2003)
  3406--15}\null.

\bibitem{LorenzBernhartHoner-zu-SiederdissenTaferFlammStadlerHofacker2011}
R.~Lorenz, S.H. Bernhart, C.~H{\"o}ner~zu Siederdissen, H.~Tafer, C.~Flamm,
  P.F. Stadler  and I.L. Hofacker,
\newblock {\em ViennaRNA package 2.0},
\newblock \doi{10.1186/1748-7188-6-26}{\rm Algorithms for Molecular Biology
  {\bf 6} (2011)}\null.



\bibitem{rissone2022stem}
P. Rissone, C.V. Bizarro, and F. Ritort,
\newblock {\em Stem--loop formation drives RNA folding in mechanical unzipping experiments},
\newblock \doi{10.1073/pnas.2025575119}{\rm  Proc. Natl. Acad. Sci. U.S.A. {\bf  119}, e2025575119 (2022)}.

\bibitem{slutsky2004diffusion}
M. Slutsky, M. Kardar and L. A. Mirny,
\newblock {\em Diffusion in correlated random potentials, with applications to DNA.},
\newblock \doi{10.1103/PhysRevE.69.061903}{\rm  Phys. Rev. E  {\bf  69}
  (2004)   061903}.
  

\bibitem{sales2003temperature}
M. Sales, J. P. Bouchaud and F. Ritort,
\newblock {\em Temperature shifts in the Sinai model: static and dynamical effects},
\newblock \doi{10.1088/0305-4470/36/3/306}{\rm  J. Phys. A t  {\bf  36}
  (2003)   665-684}.
  
\bibitem{de2015temperature}
S. De Lorenzo, M. Ribezzi-Crivellari, J. Arias-Gonzalez, S. B. Smith and F. Ritort,
\newblock {\em A temperature-jump optical trap for single-molecule manipulation.},
\newblock \doi{10.1016/j.bpj.2015.05.017}{\rm  	Biophys. J. {\bf  108}
  (2015)   2854-2864}.
  
  

\bibitem{HuguetRibezziCrivellariBizarroRitort2017}
J.M. Huguet, M. Ribezzi-Crivellari, C.V. Bizarro and F. Ritort,
\newblock {\em Derivation of nearest-neighbor DNA parameters in magnesium from single molecule experiments},
\newblock \doi{10.1093/nar/gkx1161}{\rm   Nucleic Acids Res.   {\bf  45}
  (2017)   12921-12931}\null,


\bibitem{JosepHuguetBizarroFornsSmithBustamanteRitort2010}
J.M. Huguet, C.V. Bizarro, N. Forns, S.B. Smith, C. Bustamante, and F. Ritort,
\newblock {\em Single-molecule derivation of salt dependent base-pair free energies in DNA},
\newblock \doi{10.1073/pnas.1001454107}{\rm  Proc. Natl. Acad. Sci {\bf  107}
  (2010)   15431-15436}\null,
\newblock \arxiv{arXiv:1010.1188}.  
  
  
\end{thebibliography}
\end{document}